\documentclass[12pt]{iopart}
\usepackage{iopams}
\usepackage{graphicx}

\begin{document}

\title{Exact nonlinear Bloch-state solutions for Bose-Einstein condensates in a periodic array of quantum wells}

\author{Rui Xue}
\address{Institute of Theoretical Physics and Department of
Physics, Shanxi University, Taiyuan 030006, China}

\author{Z. X. Liang}
\address{Shenyang National Laboratory for
Materials Science, Institute of Metal Research, Chinese Academy of
Sciences, Wenhua Road 72, Shenyang 110016, China}
\ead{zhxliang@gmail.com}

\author{Weidong Li}
\address{Institute of Theoretical Physics and Department of
Physics, Shanxi University, Taiyuan 030006, China}
\ead{wdli@sxu.edu.cn}

\begin{abstract}
A set of exact closed-form Bloch-state solutions to the stationary
Gross-Pitaevskii equation are obtained for a Bose-Einstein
condensate in a one-dimensional periodic array of quantum wells,
i.e. a square-well periodic potential. We use these exact solutions
to comprehensively study the Bloch band, the compressibility,
effective mass and the speed of sound as functions of both the
potential depth and interatomic interaction. According to our study,
a periodic array of quantum wells is more analytically tractable
than the sinusoidal potential and allows an easier experimental
realization than the Kr\"onig-Penney potential, therefore providing
a useful theoretical model for understanding Bose-Einstein
condensates in a periodic potential.
\end{abstract}

\maketitle

\section{Introduction}
Bose-Einstein condensates (BECs) in periodic potentials have
attracted great interest both experimentally and theoretically
during the past few years \cite{Morschrev,Blochrev}. A major reason
is that they usually exhibit phenomena typical of solid state
physics, such as the formation of energy bands \cite{Wu,Smith},
Bloch oscillations \cite{Choi,Morsch}, Landau-Zener tunneling
\cite{Loop1,Zobay,Loop2,Jona} between Bloch bands and Josephson
effects \cite{Anderson,Cataliotti}, etc. The advantage of BECs in
periodic potentials over a solid-state system is that the potential
geometry and interatomic interactions are highly controllable. Such
a BEC system can therefore serve as a quantum simulator \cite{Bloch}
to test fundamental concepts. For instance, the Bose-Hubbard model
is almost perfectly realized in the BEC field, hence enabling an
experimental study of the quantum phase transition between a
superfluid and Mott insulator \cite{Jaskch,Greiner}.

Research so far has been primarily focused on BECs in two types of
periodic potentials. The first type is the sinusoidal optical
lattice \cite {Morschrev,Blochrev}. Experimentally created by two
counter-propagating laser beams, the sinusoidal optical lattice
consists of only one single Fourier component. Most studies on BECs
in this type of potential ask for the help of numerical simulations
since analytical solutions are lacking. By contrast, the second one
is the so-called Kr\"onig-Penney potential \cite
{Li,Seaman,Danshita}. In the BEC field, the Kr\"onig-Penney
potential as shown in Refs. \cite {Li,Seaman,Danshita} is usually
referred to as a periodic delta function potential. However, in the
original work \cite{KP} and the field of condensed matter physics
\cite{Kittel}, the Kr\"onig-Penney potential is also used as the
periodic rectangular potential. To avoid the confusion, we adopt the
notion in the BEC field and refer to the Kr\"onig-Penny potential as
a periodic delta function potential. With this understanding, the
periodic rectangular potential in this paper is called as a periodic
array of quantum wells. The Kr\"onig-Penny potential admits an exact
solution in closed analytical form, leading to general expressions
that can simultaneously describe all parameter regimes.
Nevertheless, it's very difficult to realize a Kr\"onig-Penney
potential in experiments. It is therefore instructive to seek a
periodic potential that not only permits an exact solution in closed
analytical form, but also is hopeful to be realized experimentally.
The search for such type of potential is justified by the fact that
the fundamental properties of a BEC in a periodic potential should
not depend on the potential shape \cite{Seaman}. So theorists are
actually at liberty to select the form of a periodic potential for
the convenience of their study.

One such option is provided by a periodic array of quantum wells
separated by barriers \cite{Kittel}. On the experimental side, this
potential can be generated by interference of several laser beams.
Since two interference counter-propagating laser beams form a
sinusoidal potential that contains one single Fourier component, we
expect more Fourier components to be involved by using several
counter-propagating laser beams. When frequencies of these beams are
multiples of the fundamental, interference of them would result in a
periodic array of quantum wells. An experimental scheme to create
such unconventional optical lattices has recently been demonstrated
in Ref. \cite{Ritt}. On the theoretical side, it will be shown in
this paper that exact closed-form solutions exist for a periodic
array of quantum wells. In fact, such potential virtually becomes a
Kr\"onig-Penney potential, i.e., a lattice of delta functions, in
the limit when the width of barriers becomes much smaller than the
lattice period. We are therefore motivated to launch a systematic
study on a BEC in a periodic array of quantum wells.

In this article, we derive a set of exact Bloch-state solutions to
the stationary Gross-Pitaevskii equation (GPE) for a BEC in a
one-dimensional periodic array of quantum wells. All our exact
solutions, in the limit of varnishing interatomic interaction, are
reduced to their counterparts in the linear case, i.e. the Bloch
states of the stationary Schr\"odinger equation with a
one-dimensional periodic array of quantum wells. We apply these
solutions to analyze the structure of Bloch bands, the
compressibility, effective mass and the speed of sound as functions
of both potential depth and the strength of interatomic interaction.
Special emphasis is given to the behavior of the compressibility and
effective mass.

The outline of the paper is as follows. In Sec. 2, we introduce
notations and describe basic theoretical framework of our study. In
Sec. 3, the general solutions of GPE in a single quantum well are
derived in detail. In Sec. 4, we investigate the lowest Bloch band
for a BEC in a periodic array of quantum wells. A comprehensive
analysis is presented in Sec. 5 that explains the dependence of
Bloch band, the compressibility, effective mass and the speed of
sound on the potential depth and the strength of interatomic
interaction. Finally, we discuss their experimental implications
followed by a summary in Sec. 6.

\section{Mean-field theory of Bose-Einstein condensates}

We consider a BEC is tightly confined along the radial directions
and subjected to a periodic potential in x-direction. The periodic
potential $V_{pot}\left(x\right)$ is assumed to be a periodic array
of quantum wells in the form
\begin{equation}  \label{poten}
V_{pot}(x)=\sum_{n=-\infty}^{+\infty}V_{q}(x-nT),
\end{equation}
with
\begin{equation}\label{potena}
V_q\left(x\right) =\left\{
\begin{array}{lcl}
0 & \hspace{0.5cm} & 0<x\leq a, \\
sE_R & \hspace{0.5cm} & a<x\leq a+b,
\end{array}
\right.
\end{equation}
where $a$ is the well width and $b$ is the barrier width. In Eq.
(\ref{poten}), the $V_{pot}\left(x\right)$ has a periodicity of
$T=a+b $. The $s$ in Eq. (\ref{potena}) is a dimensionless parameter
that denotes the strength of the $ V_{pot}\left(x\right)$ in units
of the recoil energy $E_R=\hbar^2q_B^2/2m$, with $q_B=\pi/T$ being
the Bragg momentum.

We restrict ourselves to the case where the BEC system can be well
described by the mean-field theory. The parameter characterizing the
role of interactions in the system is $g_{3D}n$, where $g_{3D}=4\pi
\hbar^2 a_s/m$ is the two-body coupling constant and $n$ is the 3D
average density. Here $a_s>0$ is the 3D s-wave scattering length. At
the mean-field level, descriptions of a BEC system are given by the
stationary GPE (or nonlinear Schr\"{o}dinger equation). In our case,
the confinement along the radial direction is so tight that the
dynamics of the atoms in the radial direction is essentially frozen
to the ground state of the corresponding magnetic trap. As shown in
Ref. \cite{Oslshanii}, the effective coupling constant can be
deduced as $g=g_{3D}/2l_{0}^2$ with $l_0$ being the length scale of
the magnetic trap. In this limit, the stationary 3D GPE therefore
reduces to a 1D equation that reads \cite{Meret}:
\begin{equation}
\left( -\frac{\hbar^{2}}{2m}\frac{\partial^{2}}{\partial x^{2}}%
+V_{pot}\left( x\right)+gnT|\psi|^{2}\right) \psi\left( x\right)
=\mu\psi\left( x\right),  \label{g1}
\end{equation}
where $m$ is the atomic mass, $\mu$ is the chemical potential and
the order
parameter $\psi(x)$ is normalized according to $\int_{0}^{T}dx|%
\psi(x)|^2=1$.

Despite non-linearity, Eq. (\ref{g1}) permits solutions in the form
of Bloch waves \cite{Wu,sound2}
\begin{equation}  \label{Bcondition}
\psi_{k}(x)=e^{ikx}\phi_k(x),
\end{equation}
where $k$ is the Bloch wave vector and $\phi_k(x)$ is a periodic
function with the same periodicity as the $V_{pot}\left(x\right)$.
We point out that Eq. (\ref{Bcondition}) does not exhaust all
possible stationary solutions of GP Eq. (\ref{g1}) due to the
presence of the nonlinear term. Except for the Bloch-form solutions,
the GP Eq. (\ref{g1}) with a periodic potential also allows other
kinds of solutions, for example, period-doubled state solutions
\cite{Machholm}.

The GP equation (\ref{g1}), in terms of the function $\phi_k(x)$,
can be rewritten as:
\begin{equation}  \label{g3}
\left[\frac{\left(-i\hbar\partial_x+k\right)^2}{2m}+V_{pot}+gnT|\phi_k|^2
\right]\phi_k=\mu(k)\phi_k.
\end{equation}
Note that the chemical potential $\mu(k)$, which is derived from Eq.
(\ref{g3}) as
\begin{eqnarray}
\mu(k)=\int_{0}^{T}dx\phi_{k}^{*}\left[ \frac{(-i\hbar\partial_x+k)^{2}}{2m}%
+V_{pot} +gnT|\phi_{k}|^{2}\right] \phi_{k}, \label{chemical}
\end{eqnarray}
usually does not coincide with the energy $\varepsilon(k)$ of the
BEC system defined by
\begin{eqnarray}
\varepsilon(k)=\int_{0}^{T}dx\phi_{k}^{*}\left[ \frac{(-i\hbar%
\partial_x+k)^{2}}{2m}+V_{pot} +\frac{1}{2}gnT|\phi_{k}|^{2}\right] \phi_{k}.
\label{energy}
\end{eqnarray}
Comparison of Eqs. (\ref{chemical}) and (\ref{energy}) indicates
that $\varepsilon(k)$ equals $ \mu(k)$ only when interactions are
absent. Generally, $\varepsilon(k)$ and $\mu(k)$ are related to each
other through following definition  \cite{Meret,sound2}
\begin{equation}
\mu(k)=\frac{\partial\left[n\varepsilon(k)\right]}{\partial n}.
\end{equation}

Now we seek exact solutions of Eq. (\ref{g1}) by assuming following
ansatz \cite{Li,Seaman,Danshita} for the wave function $\psi(x)$
\begin{equation}  \label{ansatz}
\psi\left( x\right) =\sqrt{\rho\left( x\right) }\exp\left[
-i\Theta\left( x\right)\right],
\end{equation}
where the density function $\rho\left( x\right)$ is nonnegative and
the
phase function $\Theta\left( x\right)$ is real. Substituting Eq. (\ref%
{ansatz}) into Eq. (\ref{g1}) and re-scaling equations, we obtain
\begin{equation}
\left( \frac{\partial\rho }{\partial x}\right) ^{2}
=2\eta\rho^{3}+4\left( \mu-V_{pot} \right) \rho^{2}-\beta\rho
-4\alpha^{2} ,  \label{sita2}
\end{equation}
and
\begin{equation}
\Theta=\int dx\frac{\alpha}{\rho\left( x\right) },
\end{equation}
where the length scale is $T/\pi$, the $\eta=gnT/E_{R}$ represents
the nonlinear interaction, and $\beta$ and $\alpha$ are integral
constants.

\section{General solution in a single quantum well}
We then proceed to solve the GP Eq. (\ref{sita2}) in a single
quantum well $ V_{sig}$ defined by
\begin{equation}  \label{potsig}
V_{sig}\left(x\right) =\left\{
\begin{array}{lcl}
0 & \hspace{0.5cm} & 0<x\leq a, \\
s & \hspace{0.5cm} & a<x\leq a+b.
\end{array}
\right.
\end{equation}
As is well known, general solutions for Eq. (\ref{sita2}) with a
constant
potential can be expressed in terms of the Jacobi Elliptic functions \cite%
{1,2,6}. In our case, we derive exact solutions to Eq. (\ref{sita2})
separately in the two regions shown in Eq. ({\ref{potsig}}).

In the region $0<x\leq a$, the $V_{sig}(x)$ is zero. Hence the exact
solutions of Eq. (\ref{sita2}) have the following general form
\begin{equation}
\rho _{1}\left( x\right) =A-\left[ A-\frac{2\alpha ^{2}}{A\left(
2K^{2}+A\eta \right) }\right] $SN$^{2}\left( Kx+\delta
,n_{1}^{2}\right),   \label{solution1}
\end{equation}%
where
\begin{eqnarray}
n_{1}^{2} &=&-\frac{A}{2K^{2}}\eta +\frac{\alpha ^{2}}{AK^{2}\left(
2K^{2}+A\eta \right) }\eta ,  \nonumber \\
\beta  &=&-\frac{2\left( 2AK^{2}+A^{2}\eta \right) ^{2}+8\left(
A\eta
+K^{2}\right) \alpha ^{2}}{2AK^{2}+A^{2}\eta }, \\
\mu  &=&K^{2}+\left( A+\frac{\alpha ^{2}}{A\left( 2K^{2}+A\eta \right) }%
\right) \eta .  \nonumber
\end{eqnarray}%
In Eq. (\ref{solution1}), $SN$ is the Jacobi elliptic sine function and $%
n_{1}^{2}$ denotes the modulus whose range is restricted within
$[0,1]$. In this general solution, the free variables are the
translational scaling $K$, the translational offset $\delta $, and
the density offset $A$. In the limit of $\eta =0$, the solution
(\ref{solution1}) can be reduced to
\begin{equation}
\rho _{1}\left( x\right) =A-\left( A-\frac{\alpha
^{2}}{AK^{2}}\right) \sin ^{2}\left( Kx+\delta \right) ,
\end{equation}%
with $n_{1}=0$, $\beta =-\frac{8A^{2}K^{2}+8\alpha ^{2}}{2A}$ and
$\mu =K^{2} $.

In the region $a<x\leq a+b$, the $V_{sig}(x)$ is a constant. In this
region, Eq. (\ref{sita2}) admits two kinds of exact solutions,
depending on whether there is a node within the barrier.

The first type of solutions contains no node within the barrier and
has the form:
\begin{equation}
\rho _{2}\left( x\right) =B+\left[ B+\frac{2\alpha ^{2}}{B\left(
2Q^{2}-B\eta \right) }\right] $SC$^{2}\left( Qx+\gamma
,n_{2}^{2}\right),   \label{solution2}
\end{equation}%
with
\begin{eqnarray}
n_{2}^{2} &=&1-\left( \frac{B}{2Q^{2}}+\frac{\alpha
^{2}}{BQ^{2}\left(
2Q^{2}-B\eta \right) }\right) \eta ,  \nonumber \\
\mu  &=&s-Q^{2}+\left( B-\frac{\alpha ^{2}}{B\left( 2Q^{2}-B\eta \right) }%
\right) \eta , \\
\beta  &=&4BQ^{2}-2B^{2}\eta +\frac{\left( 8B\eta -8Q^{2}\right) \alpha ^{2}%
}{2BQ^{2}-B^{2}\eta }.  \nonumber
\end{eqnarray}%
In the limit of $\eta =0$, this solution is reduced to
\begin{equation}
\rho _{2}\left( x\right) =B+\left( B+\frac{\alpha
^{2}}{BQ^{2}}\right) \sinh ^{2}\left( Qx+\gamma \right) ,
\end{equation}%
with $n_{2}=1$, $\mu =s-Q^{2}$ and $\beta =(4B^{2}Q^{2}-4\alpha
^{2})/B$.

The second type of solutions admits only one node within the barrier
and is expressed as
\begin{equation}
\rho _{2}=-\frac{B}{8Q^{2}}+\left[ \frac{Q^{2}}{\eta }+\frac{B}{8Q^{2}}-%
\frac{Q\sqrt{BQ^{2}-16\alpha ^{2}\eta }}{\sqrt{B}\eta }\right] $NC$%
^{2}\left( Qx+\gamma ,n_{2}^{2}\right),   \label{solution3}
\end{equation}%
with
\begin{eqnarray}
n_{2}^{2} &=&\frac{1}{2}-\frac{B}{16Q^{4}}\eta
+\frac{\sqrt{BQ^{2}-16\alpha
^{2}\eta }}{2\sqrt{B}Q},  \nonumber \\
\mu  &=&s-\frac{B}{16Q^{4}}\eta -\frac{Q\sqrt{BQ^{2}-16\alpha ^{2}\eta }}{%
\sqrt{B}},  \label{b2} \\
\beta  &=&\frac{32Q^{2}\alpha
^{2}}{B}-\frac{\sqrt{B}\sqrt{BQ^{2}-16\alpha ^{2}\eta }}{2Q}.
\nonumber
\end{eqnarray}%
which again can be reduced in the limit of $\eta =0$ to:
\begin{equation}
\rho _{2}\left( x\right) =-\frac{B}{8Q^{2}}+\frac{B}{8Q^{2}}\cosh
^{2}\left( Qx+\gamma \right) ,
\end{equation}%
with $n_{2}=1$, $\mu =s-Q^{2}$ and $\beta =(B^{2}-64Q^{2}\alpha
^{2})/2B$. In Eqs. (\ref{solution2}) and(\ref{solution3}), $SC$ and
$NC$ are also the Jacobi elliptic functions with modulus $n_{2}^{2}$
and the free variables are $B$, $Q$ and $\gamma$.

Note that all above solutions, in the limit of $\eta=0$, are nothing
but the stationary solutions for the linear Schr\"odinger equation.

\section{Bloch bands and group velocity}
So far we have ignored Bloch wave condition (\ref{Bcondition}) and
solved the GP equation for a \textit{single} quantum well for
specific regions. Next we seek the \textit{global} solution to the
GP equation defined on the whole x axis that satisfies the Bloch
wave condition.

Assume that the solutions given in Eqs. (\ref{solution1}),
(\ref{solution2}) and (\ref{solution3}) respectively comprise a
segment of the complete Bloch
wave stationary solution of the GP equation with the potential given by Eq. (%
\ref{poten}). We then extend the wave function $\psi_{k}(x)$
originally defined on $\left (0, a+b\right]$ to the whole x axis and
construct the ultimate Bloch wave solution according to the Bloch
condition:
\begin{equation}  \label{Bcondition1}
\psi_k(x+T)=e^{ikT}\psi_k(x),
\end{equation}
where $k$ is the Bloch wave vector defined by
\begin{equation}
\label{Bcondition2}
k=-\Theta(x)=-\alpha\int_{0}^{T}\frac{dx}{\rho(x)}.
\end{equation}
Imposing the boundary condition that the $\psi_{k}(x)$ is continuous
at $x=a$ and using Eq. (\ref{Bcondition1}), we obtain
\begin{eqnarray}
\rho_{1}\left(  a\right)   &  =\rho_{2}\left(  a\right)  ,\label{1}\\
\partial_{x}\rho_{1}\left(  a\right)   &  =\partial_{x}\rho_{2}\left(
a\right)  ,\label{2}\\
\rho_{1}\left(  0\right)   &  =\rho_{2}\left(  T\right)  ,\label{3}\\
\partial_{x}\rho_{1}\left(  0\right)   &  =\partial_{x}\rho_{2}\left(
T\right)  . \label{4}
\end{eqnarray}
Two additional constraints are the continuity of the chemical
potential on the boundary and the normalization condition for the
wave function, i.e.
\begin{equation}
\mu_{1}=\mu_{2},   \label{6}
\end{equation}
and
\begin{equation}
\int_{0}^{T}|\psi_{k}(x)|^2dx=\int\nolimits_{0}^{a}\rho_{1}\left(
x\right) dx+\int\nolimits_{a}^{T} \rho_{2}\left( x\right)dx=1.
\label{7}
\end{equation}
In principle, Eqs. (\ref{Bcondition2}-\ref{7}) provide a complete
set of equations for us to determine the unknown parameters $A$,
$B$, $K$, $Q$, $\delta$, $\gamma$ and $\alpha$. Once these
parameters are found, we obtain the Bloch band of the periodic
system. Note that in the region $a<x\leq a+b$, the first type
solution $\rho_2(x)$ is adopted when $k\neq \pm q_B$ , while the
second type $\rho_2(x)$ is used when $k=\pm q_B$.

\begin{figure}[tbh]
\begin{center}
\rotatebox{0}{\resizebox *{9cm}{7.5cm} {\includegraphics
{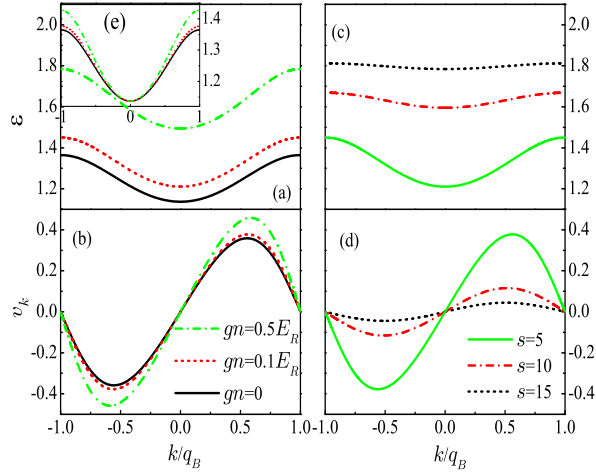}}}
\end{center}
\par
\vspace{-0.6 cm} \caption{(a) and (b) the lowest Bloch band and
group velocity with $s=5$ for $gn=0,$ $0.1E_{R}$ and $0.5E_{R}$; (c)
and (d) the lowest Bloch band, group velocity with $s=5$, $10$, $15$
when $gn=0.1E_{R}$. (e) in order to compare
changes of the energy band, we vertically translate the Bloch band of $%
gn=0.1E_{R}$ and $0.5E_{R}$ to the initial position of $gn=0$.}
\label{fig1}
\end{figure}

In Fig. \ref{fig1}, the lowest Bloch band and the corresponding
group velocity as a function of quasi momentum $k$ are presented for
various potential depth $s$ and interatomic interactions $gn$.
Quantum well width $a=0.6$, and barrier width $b=0.4$ are used
throughout this paper. Here, the group velocity is defined by
$v_{k}=\partial\varepsilon/\partial k$. The states with $k=0$ and
$k=\pm q_B$ in Fig. \ref{fig1} respectively correspond to the
stationary condensates at the bottom and top of the lowest Bloch
band. The state with $k\neq 0$ and $k \neq \pm q_B$ in Fig.
\ref{fig1}, on the other hand, describe a condensate where all atoms
occupy the same single-particle wave function and move together in
the periodic potential with a constant current $nv_{k}$.

Furthermore, Figs. \ref{fig1} (a), (e) and (b) demonstrate that when
the potential depth $s$ is fixed, the interatomic interactions
affect the group velocity more conspicuously than the Bloch band.
Yet for given interatomic interaction $gn$, Fig. \ref{fig1} (c) and
(d) show that the Bloch band becomes more and more flatter with
increasing potential depth. Eventually, when the potential wells are
sufficiently deep, the condensate becomes so localized in each
quantum well that an adequate description can be obtained by
directly using the tight-binding model \cite{Meret}.

\section{Compressibility, Effective mass and sound speed}

Now we apply our exact solutions to study the compressibility, the
effective mass and the sound speed of a BEC in a periodic array of
quantum wells.

We start by calculating the compressibility $\kappa$. In
thermodynamics, $\kappa$ is defined as the relative volume change of
a fluid or solid with respect to a pressure (or mean stress)
variation. In our case, the compressibility $\kappa$ is given
by\cite{Meret,sound2}
\begin{equation}
\kappa^{-1}=n\frac{\partial\mu}{\partial n}.   \label{compress}
\end{equation}
For a BEC system with repulsive interatomic interaction, the
periodic potential traps atoms and enhances the repulsion. A reduced
compressibility $\kappa$ is therefore expected. We illustrate this
point in detail in the following.

\begin{figure}[tbh]
\begin{center}
\rotatebox{0}{\resizebox *{9cm}{7.5cm} {\includegraphics
{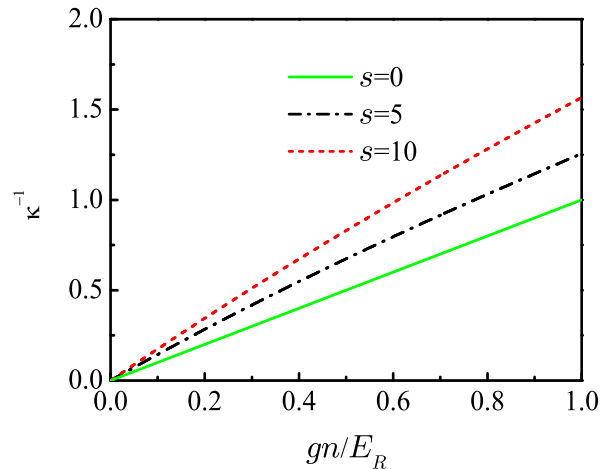}}}
\end{center}
\par
\vspace{-0.5 in}
\caption{Inverse compressibility $\protect\kappa^{-1}$ as a function of $%
gn/E_{R}$ for $s=0$, $5$, $10.$} \label{Fig2}
\end{figure}

In the uniform case of $s=0$, the chemical potential is linearly
dependent on the density expressed by $\mu=gn$. Thus
$\kappa^{-1}=gn$ is proportional to the density. When $s \neq 0$, we
substitute the general expression of $\mu$ given by Eq.
({\ref{chemical}}) into Eq. ({ \ref{compress}}), and obtain the
$\kappa^{-1}$ for a BEC system in a periodic array of quantum wells.
The calculated $\kappa^{-1}$ is plotted in Fig. \ref{Fig2} as a
function of the interatomic interaction $gn$ for different $s$. The
figure demonstrates that the $\kappa^{-1}$ increases with $s$,
typical of a wave function localized at the bottom of each quantum
wells. Compared to the uniform case, the $\kappa^{-1}$ increases
linearly only for small $gn/E_R$. Whereas for large $gn/E_R$, the
growth of $\kappa^{-1}$ develops a nonlinear dependence on $gn/E_R$.

We now give an analytical explanation to the behavior of
$\kappa^{-1}$ shown in Fig. \ref{Fig2}. Assume that $\kappa^{-1}$ is
related to $s$ by the following expression \cite{Meret} when
$gn/E_R$ is small
\begin{equation}
\kappa^{-1}=\widetilde{g}\left( s\right) n ,  \label{effec}
\end{equation}
where
\begin{equation}
\mu=\mu_{gn=0}+\widetilde{g}\left( s\right) n,
\end{equation}
in which $\mu_{gn=0}$ depends on the potential depth, but not on
density. The quantity $\widetilde{g}\left( s\right)$ in Eq.
(\ref{effec}) acts as an effective coupling constant. In case Eq.
(\ref{effec}) is valid,  the compressibility of a BEC in a periodic
array of quantum wells with $g$ is virtually transformed to the
compressibility of a uniform BEC with the $\widetilde{g}(s)$. Thus
by simply replacing $g$ by $\widetilde{g}(s)$, we can view our
system as if there is no periodic potential \cite{Meret} as far as
the compressibility is concerned.

To obtain the form of $\widetilde{g}\left( s\right)$, we substitute
Eq. (\ref{chemical}) into Eq.  (\ref{compress}) yielding
\cite{Meret}

\begin{equation}
\kappa^{-1}=n\frac{\partial\mu}{\partial n}=gn\int_{0}^{T}\phi^{4}_{\eta=0}%
\left( x\right) dx ,  \label{g2}
\end{equation}
where $\phi_{\eta=0}$ is the ground state solution of Eq (\ref{g3}) for $%
\eta=0$. Comparison of Eq. (\ref{g2}) with Eq. (\ref{effec}) gives
\begin{equation}
\widetilde{g}=g\int_{0}^{T}\phi^{4}_{\eta=0}\left( x\right) dx,
\label{coupl}
\end{equation}
which in our formulation has the following form:
\begin{equation}
\widetilde{g}=g\int_{0}^{a}\left. \rho_{1}^{2}\left( x\right)
\right\vert _{\eta=0}dx+g\int_{a}^{T}\left. \rho_{2}^{2}\left(
x\right) \right\vert _{\eta=0}dx,
\end{equation}
where $\rho_{1}\left( x\right) $ and $\rho_{2}\left( x\right) $ are
solutions respectively in well and barrier.

\begin{figure}[tbht]
\begin{center}
\rotatebox{0}{\resizebox *{9cm}{7.5cm} {\includegraphics
{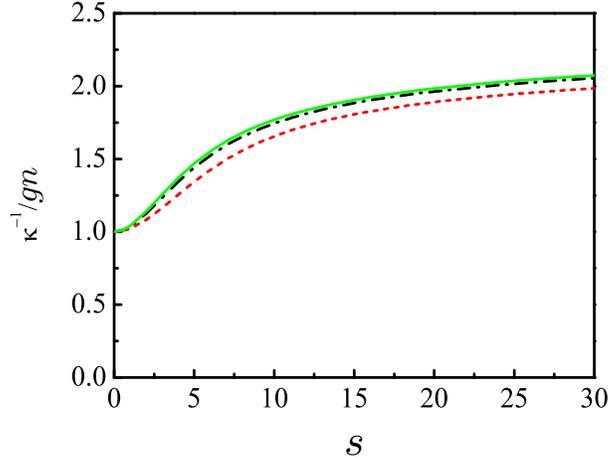}}}
\end{center}
\par
\vspace{-0.5 in} \caption{$\protect\kappa^{-1}/gn$ for $gn=0.1E_{R}$
(dashed-dotted line) and $gn=0.5E_{R}$ (short dashed line) as a
function of the potential depth $s$, comparing with the effective
coupling constant $\widetilde{g}/g$ (solid line).} \label{Fig3}
\end{figure}

We plot the $\kappa^{-1}/gn$ as a function of the potential depth $s$ for $%
gn=0.1E_{R}$ and $gn=0.5E_{R}$ in Fig. \ref{Fig3}. To compare with
the
behavior of the effective coupling constant $\widetilde{g}$ defined by Eq. (%
\ref{coupl}), the function of $\widetilde{g}/g$ with $s$ is also
plotted. Fig. {\ref{Fig3}} shows that the linear dependence of
$\kappa^{-1}$ on $gn$ breaks down. However, with the increasing of
$s$, the law of $\kappa^{-1}=\widetilde{g}n$ becomes to be
applicable.


\begin{figure}[tbh]
\begin{center}
\rotatebox{0}{\resizebox *{9cm}{7.5cm} {\includegraphics
{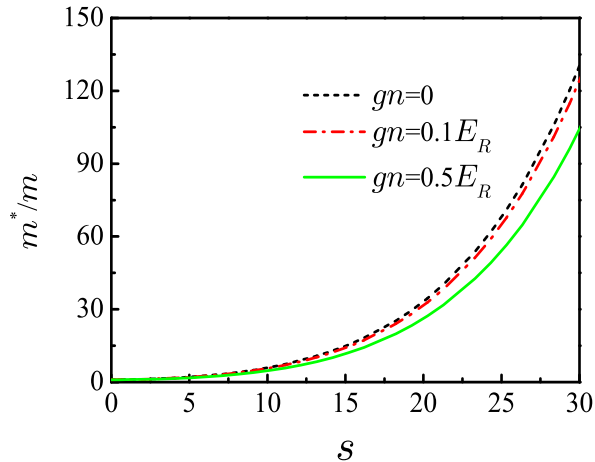}}}
\end{center}
\par
\vspace{-0.5 in}
\caption{Effective mass as a function of potential depth $s$ for $gn=0$, $%
gn=0.1E_{R}$ and $gn=0.5E_{R}.$} \label{Fig4}
\end{figure}

We then consider the effective mass. A BEC trapped in a periodic
potential can be approximately described by a uniform gas of atoms
each having an effective mass $m^{\ast}$ defined by
\cite{Meret,sound2} :
\begin{equation}
\frac{1}{m^{\ast}}=\left. \frac{\partial^{2}\varepsilon(k)}{\partial
k^{2} } \right\vert _{k=0}.
\end{equation}
The dependence of effective mass $m^{\ast}\left( k=0\right)$ on the
potential depth $s$ for $gn=0$, $gn=0.1E_{R}$ and $gn=0.5E_{R}$ is
demonstrated in Fig. \ref{Fig4}. According to Fig. \ref{Fig4}, when
$s=0$, the effective mass $m^{*}$ is reduced to the bare mass $m$.
Whereas when $s$ increases, for example, to $s=30$, the $m^{*}$
becomes two orders larger in magnitude than the $m$. This increase
of $m^*$ with $s$ can be explained by the slow-down of the particles
during their tunneling through the barriers. Fig. \ref{Fig4} also
demonstrates that the the $m^{*}$ effectively decreases with
increasing interactions. This is because repulsion, contrary to the
lattice potential that serves as a trap, tends to increase the width
of the wave function which favors tunneling. This is the so-called
screening effect of the nonlinearity \cite{Choi}.

As is emphasized in Ref. \cite{Meret}, the $m^{*}$ is determined by
the tunneling properties of the system, thereby exponentially
sensible to the behavior of wave function within the barriers. Thus
any small change in the wave function will significantly affect the
value of $m^{*}$. As a result, the conventional Gaussian
approximation \cite{Meret} in the tight-binding limit can not be
employed to calculate the $m^{*}$. In this aspect, a periodic array
of quantum wells as a solvable model, provides a better choice than
the sinusoidal potential in studying the $m^{*}$ of a BEC in a
periodic potential.

Finally, we proceed to study the sound speed. Sound is a propagation
of small density fluctuations inside a system
\cite{Meret,sound2,Zaremba,sound1,sound3}. The key point in studying
sound is to find the sound speed. The speed of sound is important
for two simple reasons: (i) it is a basic physical parameter that
tells how fast the sound propagates in the system, and (ii) it is
intimately related to superfluidity according to Landau's theory of
superfluid. Because of these, the sound propagation and its speed
were one of the first things that have been studied by
experimentalists on a BEC since its first realization in 1995
\cite{Andrews,Raman}.

The first step to derive sound speed in a BEC is to find the ground
state since it acts as a media for the sound propagation. Next, one
determines the sound speed by perturbing the ground state.
Traditionally, there are two equivalent definitions for the sound
speed \cite{sound2}. In the first definition, sound is regarded as a
long wavelength response of a system to the perturbations. Sound
speed can be extracted from the lowest Bogoliubov excitation energy,
which is characterized by the linear phonon dispersion with a finite
slope. We emphasize that the physical meaning underlying the
Bogoliubov spectrum is very different form that of the Bloch bands
discussed in Fig. 1. The Bloch bands refer to states which involve a
motion of the whole condensate through the periodic potential.
However, the Bogoliubov spectrum describes small perturbations which
involve only a small portion of atoms. The non-perturbed condensate
acts as a medium through which the perturbed portion is moving. In
other words, the Bloch band gives the energy per particle of the
current states. Being multiplied by $N$, the Bloch band energies
obviously exceed the energies of the Bogoliubov excitations. In the
second definition, the BEC system is viewed as a hydrodynamical
system. Accordingly, the sound speed in a BEC assumes following
standard expression \cite{Meret,sound2,sound1,sound3}
\begin{equation}
v_{sound}=\frac{1}{\sqrt{\kappa m^{\ast}}} .  \label{velocity}
\end{equation}
Here we adopt the second definition of sound speed in Eq.
(\ref{velocity}) in following calculations, using our previous
derivation of the compressibility and effective mass.

\begin{figure}[tbh]
\begin{center}
\rotatebox{0}{\resizebox *{9cm}{7.5cm} {\includegraphics
{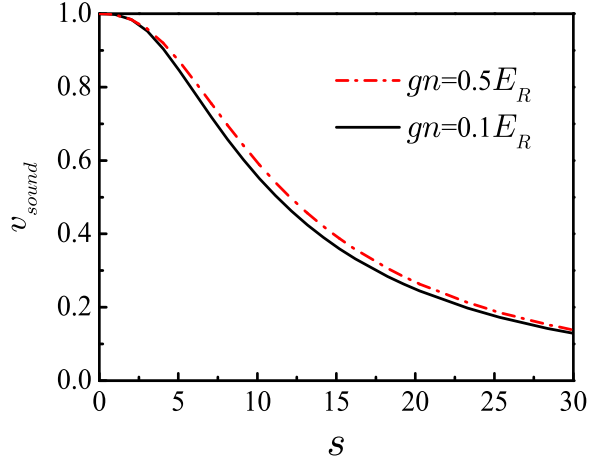}}}
\end{center}
\par
\vspace{-0.5 in} \caption{Sound speed as a function of the potential
depth $s$ divided by the
sound speed in the absence of lattice ($s=0$) for $gn=0.1E_{R}$ and $%
gn=0.5E_{R}.$} \label{Fig5}
\end{figure}

The calculation of sound speed as a function of the potential depth
$s$ is plotted in Fig. \ref{Fig5} for $gn=0.1E_{R}$ and
$gn=0.5E_{R}$.
The figure demonstrates that sound velocity decreases when the
potential depth is increased. This can be explained by the
competition between the slowly decreasing $\kappa$ and the
increasing $m^*$ when lattice depth is increased.

\section{Conclusion}

In typical experiments to date, the relevant parameters are usually
chosen as follows: the interatomic interaction $gn$ ranges from
$0.02E_R$ to $1E_R$ \cite{Morschrev,Blochrev}; the depth of the
periodic potential $s$ can be adjusted from $0E_R$ to $ 12E_R$
\cite{Greiner}, whereas the BEC system is still kept in the
superfluid state. In particular, for a one-dimensional periodic
potential, the transition to the insulator phase is expected to
happen for very deep lattice. Thus there is a broad range of
potential depths where the gas can be described as a fully coherent
system within the framework of the mean field GPE. Hence the range
of parameters in our model fit well in the current experimental
conditions. Further more, a periodic array of quantum wells could be
experimentally generated by the interference of serval
two-counter-propagating laser beams \cite{Ritt}. However, we would
like to point out that our study is based on GPE. In this mean-field
theory, all quantum fluctuations and temperature effect are ignored.
Thus in order to study the effects of temperature or fluctuations,
one has to use other theories \cite{Petrov}, especially near the
transition point of superfluid and Mott insulator.

In this paper, we obtain a set of exact closed-form Bloch-state
solutions to the stationary GPE for a BEC in a one-dimensional
periodic array of quantum wells. These solutions are applied to
calculate the Bloch band, the compressibility, effective mass and
speed of sound as functions of the potential depth and the
interatomic interaction. As a result, this type of periodic
potential provides a useful model for further understanding of BECs.

\section{Acknowledgments}

We thank Biao Wu and Ying Hu for helpful discussion. R.X. and W.D.L.
are supported by the NSF of China Grants No.10674087, 973 program
(Nos. 2006CB921603, 2008CB317103), the NSF of Shanxi Province (Nos.
200611004), NCET (NCET-06-0259). Z.X.L. is supported by the IMR
SYNL-T.S. K$\hat{e}$ Research Fellowship.

\end{document}